\documentclass[11pt,a4paper]{article} 
\pdfoutput=1
\usepackage{jcappub}
\usepackage{subfigure}
\newcommand{\eq}[1]{\begin{equation}#1\end{equation}}
\newcommand{\eqa}[1]{\begin{eqnarray}#1\end{eqnarray}}
\newcommand{\secs}[1]{\section{#1\label{sec-#1}}}
\newcommand{\ssecs}[1]{\subsection{#1\label{ssec-#1}}}

\newcommand{\fig}[4]{\begin{figure}[#4tbp]\centering\includegraphics[width=#3\textwidth]{Graph-#1.pdf}\caption{#2}\label{fig-#1}\end{figure}}
\newcommand{\figa}[3]{\begin{figure}[#3tbp]\centering #1\caption{#2}\end{figure}}
\newcommand{\figi}[3]{\subfigure[#2]{\includegraphics[width=#3\textwidth]{Graph-#1.pdf}\label{fig-#1}}}

\newcommand{\refeq}[1]{Eq.\ (\ref{eq-#1})}
\newcommand{\refsec}[1]{section \ref{sec-#1}}
\newcommand{\refssec}[1]{section \ref{ssec-#1}}

\newcommand{\refig}[1]{Fig. \ref{fig-#1}}

\newcommand{\Refeq}[1]{Eq.\ (\ref{eq-#1})}

\newcommand{\subs}[1]{_\mathrm{#1}}

\newcommand{\dd}[1]{\mathrm{d}#1}
\newcommand{\mpl}{M\subs{p}}
\newcommand{\cL}{\mathcal{L}}
\newcommand{\hf}{\frac{1}{2}}
\newcommand{\cm}[1]{}
\def\defaultfigsize{0.7}
\def\vev{VEV\ }
\newcommand{\epa}{\epsilon_\phi}
\newcommand{\epb}{\epsilon_\psi}
\newcommand{\etaa}{\eta_{\phi\phi}}
\newcommand{\etab}{\eta_{\phi\psi}}

\newcommand{\pp}{{iso}}
\newcommand{\pe}{{cur}}
\newcommand{\phib}{\delta\phi_\pe}
\newcommand{\phip}{\delta\phi_\pp}
\newcommand{\psib}{\delta\psi_\pe}
\newcommand{\psip}{\delta\psi_\pp}
\newcommand{\fNL}{f\subs{NL}}

\newcommand{\tNL}{\tau\subs{NL}}
\newcommand{\maP}{\mathcal{P}}
\newcommand{\hatt}[1]{\hat{\mathbf{#1}}}
\newcommand{\gev}{\mathrm{\ GeV}}

\begin{document}
\title{Separable and non-separable multi-field inflation and large non-Gaussianity}
\author[a,b]{Anupam Mazumdar}
\author[a]{and Lingfei Wang}
\affiliation[a]{Consortium for Fundamental Physics, Physics Department, Lancaster University, LA1 4YB, UK}
\affiliation[b]{Niels Bohr Institute, Copenhagen University, Blegdamsvej-17, Denmark}
\abstract{In this paper we provide a general framework based on $\delta N$ formalism to estimate the cosmological observables pertaining to the cosmic microwave background radiation for non-separable potentials, and for generic \emph{end of inflation} boundary conditions. We provide analytical and numerical solutions to the relevant observables by decomposing the cosmological perturbations along the curvature and the isocurvature directions, \emph{instead of adiabatic and entropy directions}.  We then study under what conditions large bi-spectrum and tri-spectrum can be generated through phase transition which ends inflation. In an illustrative example, we show that large $f_{NL}\sim {\cal O}(80)$ and $\tau_{NL}\sim {\cal O}(20000)$ can be obtained for the case of separable and non-separable inflationary potentials.}
\maketitle

\secs{Introduction}
The primordial inflation is a well tested paradigm for the early universe~\cite{Komatsu:2010fb}, which is responsible for creating the perturbations and the matter in the universe, for recent reviews see Refs.~\cite{Mazumdar:2010sa,Mazumdar:2011zd}. Typically inflation can happen via slow rolling of a single scalar field on a smooth potential, which has unique predictions for the  cosmic microwave background radiation~\cite{Mukhanov:1990me}. The induced perturbations are generically random Gaussian fluctuations with almost scale invariant spectrum with a small tilt which indicates that inflation must come to an end in our patch of the universe. The fluctuations are known to be adiabatic perturbations which sources directly the curvature perturbations relevant for the structure formation. 

However, if there were two or more scalar fields rolling simultaneously on top of the potentials, there would be more sources for the curvature perturbations. It has already been known for a while that a relative perturbation from other fields, sometimes known as the entropy perturbations can also source the curvature perturbations along with the collective adiabatic perturbations during inflation~\cite{Kodama:1985bj,Mukhanov:1990me,Gordon:2000hv}. A non-negligible coupling between the fields will typically curve the inflationary trajectory and the perturbations can be projected tangential to the trajectory (adiabatic perturbations),  and perpendicular to the trajectory (entropy perturbations)~\cite{Gordon:2000hv}.

In order to track all the components of the perturbations, we base our formalism on $\delta N$ approximation (where $N$ being the number of e-foldings) and separate universe approach~\cite{Sasaki:1995aw,Sasaki:1998ug,Wands:2000dp,Sasaki:2007ay}. The $\delta N$ formalism has been employed successfully for single and multi-field models of inflation~\cite{Liddle:1998jc}. Attempts to deal with the simplest forms of two-field potential --- the separable potential by sum and product --- have been carried out in  Refs.~\cite{GarciaBellido:1995qq, Vernizzi:2006ve,Choi:2007su,Huang:2010cy} with the $\delta N$ formalism proposed in Refs.~\cite{Sasaki:1995aw,Lyth:2005fi}. 

In Ref.~\cite{GarciaBellido:1995qq} when dealing with the separable potentials by product, the authors have suggested decomposing the perturbations into the adiabatic and entropy directions. Such a choice of decomposition has been deployed by many follow-up papers~\cite{Vernizzi:2006ve,Choi:2007su,Huang:2010cy}.  The final formulae for the relevant observables are usually complicated and indirect and usually the adiabatic limit is taken, see Refs.~\cite{Vernizzi:2006ve,Elliston:2011dr,Elliston:2011et}. There have been attempts recently on predicting cosmological observables using moment transport equations, see Ref.~\cite{Mulryne:2010rp,Mulryne:2009kh,Frazer:2011br,Seery:2012vj}. Their results are still complicated enough and no analytical expressions are available for non-separable potentials.

In this paper we decompose field perturbations into curvature and isocurvature perturbations, which are defined differently from \emph{adiabatic and entropy modes}. This allows us to present analytical solutions for separable and non-separable inflaton potentials. Our formalism provides straightforward formulae for separable potentials, and the transparency in implementing the numerics for the non-separable inflaton potentials. Furthermore, it can be employed to study the bi-spectrum and tri-spectrum during inflation. 

It is well known that single field models of slow roll inflation do not generate large non-Gaussianity~\cite{Maldacena:2002vr}. One requires a significant contribution from entropy perturbations arising from multi fields, and the conversion from entropy into curvature perturbations. The latter needs to be non-adiabatic in nature, for instance the end of inflation must happen via phase transition.

In order to generate large non-Gaussianity, the requirements for a non-trivial boundary condition and conversion of entropy perturbations were first stressed in the context of preheating~\cite{Enqvist:2004ey,Enqvist:2005qu,Jokinen:2005by}. The preheating is a phenomenon of non-adiabatic particle creation, after the end of inflation when the inflaton is coherently oscillating, for a review see~\cite{Allahverdi:2010xz}. It was shown that during preheating one can generate significant non-Gaussianity from the conversion of entropy perturbations~\cite{Jokinen:2005by}.

One can realize a very similar scenario during multi-field inflation, where the end of inflation happens not by the violation of slow roll conditions but by phase transition due to a waterfall field, as shown in Refs.~\cite{Sasaki:2008uc}, in the context of multi-brid inflation. In these models the end of inflation boundary and the inflaton trajectory intersects at an acute angle in the field space which converts the entropy perturbations into curvature perturbations at the end of inflation, and is also responsible for enhancing the observed non-Gaussianity. In this paper we will provide two examples: one with a separable potential to demonstrate how it works, and one with non-separable potential to show its capability, both of which acquire large non-Gaussianities.

Since we base the formalism on $\delta N$ approximation, it still inherits \emph{systematic errors} coming from $\delta N$ formalism. Such errors typically come from neglecting the interference between different modes and taking field perturbations as uncorrelated Gaussian at the Hubble exit~\cite{Seery:2005gb}. They may also come from the assumption that modes freeze once they go super-Hubble~\cite{Nalson:2011gc}. We will not discuss the significance of these errors as they are beyond the scope of this paper.

\secs{Formalism}
Let us first decompose a generic perturbation into perturbations along the adiabatic and entropy directions as shown in \refig{Trajectory}. The adiabatic perturbation is converted into the classical curvature perturbation at the Hubble exit, while the entropy perturbation can still be converted into the curvature perturbation after that. For instance, in hybrid models there is a possibility of converting entropy into curvature perturbation at the end of inflation due to phase transition, see~\cite{Linde:1991km,Naruko:2008sq,Sasaki:2008uc}. For a curved trajectory entropy perturbation can also be converted into curvature perturbation during inflation after the Hubble exit. For this reason we \emph{do not} take the isocurvature direction at the Hubble exit the same as the entropy direction. Instead, we define the isocurvature direction as the one along which field perturbations do not generate any curvature perturbations, i.e.\ on a uniform-$N$ hypersurface. We take the convention of $N$ as the remaining e-foldings till a uniform energy density hypersurface after which the conversion from entropy to curvature perturbation is negligible. This gives
\eq{\dd N=-H\dd t,\label{eq-Conv-dN}}
where $H\equiv\dd\ln a/\dd t$ is the Hubble expansion rate and $a$ is the scale factor. For simplicity we will always assume the slow roll conditions.

\fig{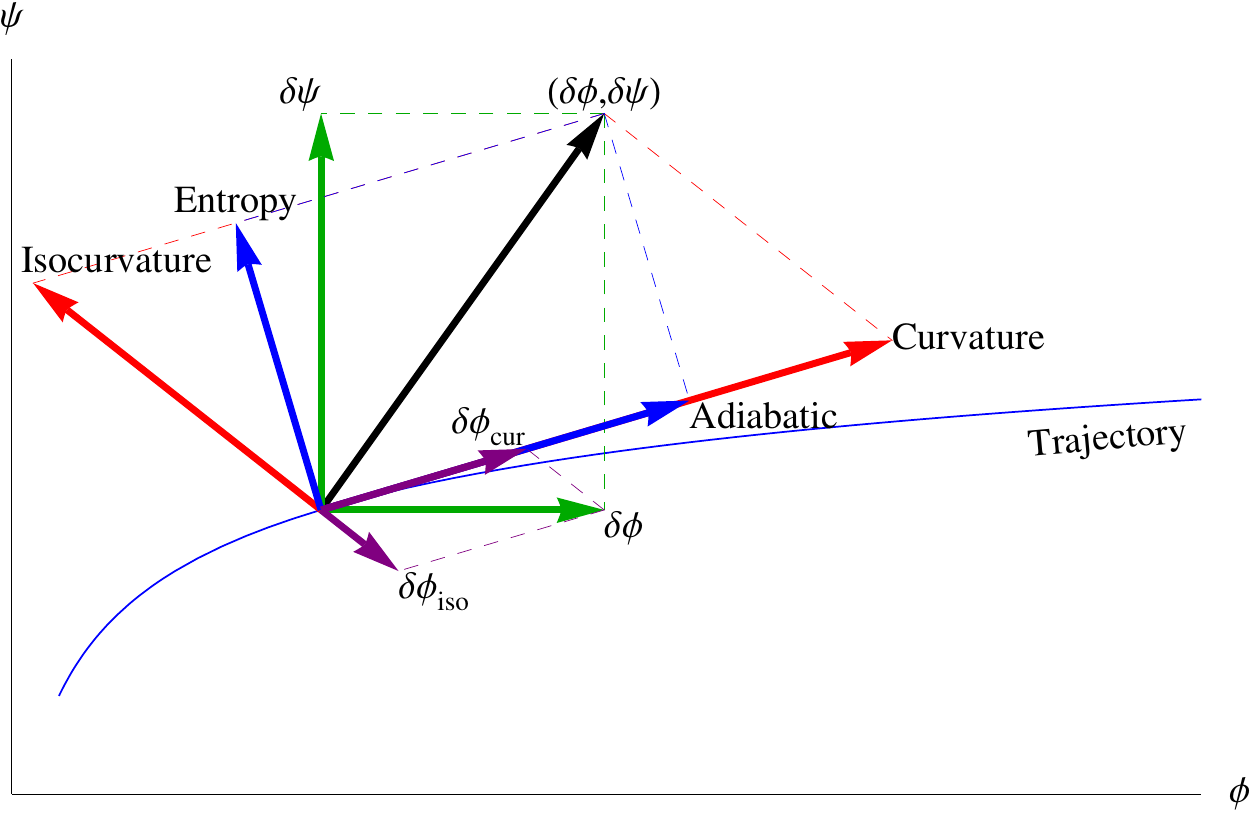}{This is a schematic figure of trajectories and perturbation decompositions in the field space. The blue curve is a general trajectory, on which a general perturbation $(\delta\phi,\delta\psi)$ at a point is shown by the black arrow. It can be decomposed to perturbations of fields $\delta\phi$ and $\delta\psi$ shown by green arrows, or to adiabatic and entropy perturbations in blue arrows, or to curvature and isocurvature perturbations in red arrows. $\delta\phi$ is taken as an example shown in purple that any perturbation including field perturbations can be decomposed to a curvature component ``$\pe$'' and an isocurvature component ``$\pp$''.}{0.77}{}
Therefore, in this paper we will decompose the field perturbations into curvature and isocurvature instead of adiabatic and entropy perturbations. The curvature direction is always along the trajectory of inflation, so it is actually the same as the adiabatic direction. Since their other respective  components have different directions, the decomposition leads to different amplitudes for adiabatic and curvature perturbations, as demonstrated in \refig{Trajectory}. This difference in amplitudes is exactly the amount of curvature perturbation converted from the entropy perturbation after the Hubble exit.

We would also like to mention here that the decomposition into curvature and isocurvature perturbations is only initially known at the end of inflation hypersurface. The decompositions during inflation should then be calculated through the method that is followed. Such calculations are necessary because all the information needed for cosmological observables is encoded in the decomposition, see \refssec{Cosmological observables and non-Gaussianity}.

\ssecs{Conventions}
Let us  consider a general two-field slow roll inflation with the Lagrangian density
\eq{\cL=-\hf\partial^\mu\phi\partial_\mu\phi-\hf\partial^\mu\psi\partial_\mu\psi-V(\phi,\psi),\label{eq-Conv-Lagrangian}}
where $\phi$ and $\psi$ are the two slow roll scalar fields whose perturbation distributions are considered independently 
Gaussian, and $V(\phi,\psi)$ is their potential. For convenience, we use a slightly different convention for the slow roll parameters
\eqa{
\epa&\equiv&\frac{\mpl V_{,\phi}}{4\sqrt\pi V},\\
\etaa&\equiv&\frac{\mpl^2 V_{,\phi\phi}}{8\pi V},\\
\etab&\equiv&\frac{\mpl^2 V_{,\phi\psi}}{8\pi V},\\
\epsilon^2&\equiv&\epa^2+\epb^2.}
Here $\mpl\equiv1/\sqrt G$ is the the Planck mass, and the subscripts $\phi$ and $\psi$ after ``,'' denote derivatives with respect to the fields. For simplicity, unless expressions for $\phi$ and $\psi$ are asymmetric, we only give the equations for $\phi$ here and onwards. 

As long as the quantum effect does not overwhelm the classical slow roll, we can define $C(\phi,\psi)$ as the variable to track the trajectory whose exact expression can be found in \refeq{Boun-Cdef}. In fact it can be chosen arbitrarily as long as it is a unique function of the trajectory, at least locally. It should also be defined as such that the infinitesimal isocurvature perturbations have nonzero $\delta C(\phi,\psi)$ and vanishing $\delta N$, while the curvature perturbations have non-vanishing $\delta N$ and vanishing $\delta C(\phi,\psi)$, which indicate $\dd N/\dd C=\dd C/\dd N=0$. As a result of this the curvature and isocurvature perturbations \emph{never} transfer into each other.

Therefore we have a two-dimensional phase space during inflation, represented by parameters either $(\phi,\psi)$ or $(N,C)$. Either parameter set can thus be expressed as a unique function of the other set, such as $N(\phi,\psi)$ and $\phi(N,C)$. For a specific trajectory, $C$ is a constant, so we may omit it sometimes and write $\phi(N)$ only. For a general partial derivative, we will use the variable explicitly in the subscript after ``,'' \footnote{Primes will denote the partial derivatives w.r.t $N$ while keeping $C$ constant but regarding $\phi$ and $\psi$ as functions of $N$, and dots will denote taking partial derivatives w.r.t $C$ keeping $N$ constant in the same way.}.

\ssecs{Background and first order perturbations}
According to $\delta N$ formalism, see~Refs.~\cite{Sasaki:1995aw, Wands:2000dp}, on large scales calculating the power spectrum of curvature perturbation is equivalent to calculating the power spectrum of $\delta N$ in separate universes on the initially flat hypersurface at the Hubble exit. The uncorrelated Gaussian perturbations $(\delta\phi,\delta\psi)$ of $(\phi,\psi)$, then generate the perturbations in the e-foldings:
\eq{\delta N=N_{,\phi}\delta\phi+N_{,\psi}\delta\psi+\frac{1}{2}\Bigl(N_{,\phi\phi}\delta\phi^2+N_{,\psi\psi}\delta\psi^2\Bigr)+N_{,\phi\psi}\delta\phi\delta\psi+\mathrm{higher\ orders}.\label{eq-Calc-deltaN}}
The power spectra of $\delta\phi$ and $\delta\psi$ take the same value:
\eq{\maP_{\delta\phi}=\maP_{\delta\psi}=\frac{H^2}{4\pi^2}.
\label{eq-Calc-phipsi}}

Now consider the period when the slow roll conditions hold for both the fields. 
The equation of motion for $\phi$ then becomes:
\eq{\frac{\dd^2\phi}{\dd t^2}+3H\frac{\dd\phi}{\dd t}+V_{,\phi}=0.}
After applying the slow roll approximations and \refeq{Conv-dN}, it reduces to
\eq{\dd\phi=\frac{\epa\mpl}{2\sqrt\pi}\dd N.\label{eq-Calc-phiprime}}
In a separate universe approach where super-Hubble perturbations are smoothened to a locally homogeneous perturbation $(\delta\phi,\delta\psi)$, the first order perturbation from \refeq{Calc-phiprime} can be written as:
\eq{\dd\delta\phi=\frac{\epa\mpl}{2\sqrt\pi}\dd\delta N+\frac{\mpl}{2\sqrt\pi}\delta\epa\dd N.\label{eq-Calc-dphiprime}}
Since our aim is to calculate $N_{,\phi}$, we take $\delta\phi$ and $\delta\psi$ to be infinitesimal. As shown in \refig{Trajectory} in purple, we decompose $\delta\phi$ in two directions:
\begin{itemize}
\item{Curvature direction: along the trajectory defined as $\phib$, and}
\item{Isocurvature direction: along the uniform-$N$ hypersurface defined as $\phip$.}
\end{itemize}

According to this definition, their separate equations from \refeq{Calc-dphiprime} are
\eqa{
\dd\phib&=&\frac{\mpl}{2\sqrt\pi}\epa\dd\delta N,\label{eq-Calc-phib}\\
\dd\phip&=&\frac{\mpl}{2\sqrt\pi}\delta\epa\dd N\nonumber\\
&=&\bigl((\etaa-2\epa^2)\phip+(\etab-2\epa\epb)\psip\bigr)\dd N.\label{eq-Calc-phip}}
Note that these two equations have different meaning. \Refeq{Calc-phip} tells us how $\phip$ evolves as the universe is evolving, while \Refeq{Calc-phib} tells us how many $\delta N$ would be generated (linearly) by the perturbations $\phib$. So we can remove the $\dd$'s in \refeq{Calc-phib} and rewrite it as
\eq{\phib=\frac{\mpl}{2\sqrt\pi}\epa\delta N.\label{eq-Calc-phib2}}
Similar expressions can be derived for the $\psi$ field. The \refeq{Calc-phip} together with  the $\psi$ version becomes a set of differential equations for $\phip$ and $\psip$, provided with the background trajectory, $C$. These differential equations have the set of solutions given by:
\eq{\phip=\frac{\partial\phi_\pp}{\partial C}\delta C=f_\phi(N,C)\delta C,\label{eq-Calc-phipsol}}
where $f_\phi$ is some function of $N$ and $C$, and $\delta C$ is the deviation from the original trajectory which stays constant as the separate universe evolves.

Since the curvature and isocurvature directions are not necessarily perpendicular, solving $N_{,\phi}$ becomes more complicated. First of all, it is defined as
\eq{N_{,\phi}\equiv\frac{\partial N}{\partial \phi}=\left.\frac{\delta N}{\delta\phi}\right|_{\delta\psi=0}.}
Constructing the $\delta\psi=\psib+\psip=0$ direction from the $\psi$ version of \refeq{Calc-phib2} and \refeq{Calc-phipsol} yields:
\eq{\delta C=-\frac{\epb\mpl}{2\sqrt\pi f_\psi}\delta N.\label{eq-Calc-deltaC}}
This is followed by
\eq{\delta\phi=\phib+\phip=\frac{\epa f_\psi-\epb f_\phi}{2\sqrt\pi f_\psi}\mpl\delta N,}
which provides
\eq{N_{,\phi}=\frac{2\sqrt\pi}{\mpl}\frac{f_\psi}{\epa f_\psi-\epb f_\phi}.\label{eq-Calc-Nphi1}}
Since $\epsilon_\phi$ and $f_\phi$ are both functions of $N$ for a given trajectory, we solve $N_{,\phi}$ by taking the derivate w.r.t $N$ on both sides of \refeq{Calc-Nphi1}, while $\phi$ and $\psi$ are regarded as functions of $N$. After some calculation and substituting \refeq{Calc-phip}, we reduce it to
\eq{N_{,\phi}'=u_\phi N_{,\phi}+v_\phi,\label{eq-Calc-Nphip}}
where the prime denotes taking derivative w.r.t $N$ while fixing $C$, and 
\eqa{
u_\phi&\equiv&\frac{\epa}{\epb}\etab-\etaa,\\
v_\phi&\equiv&\frac{2\sqrt\pi}{\mpl}\left(2\epa-\frac{\etab}{\epb}\right).}
The above differential equation has an exact analytic solution for $N_{,\phi}$ as a function of $N$ (for a non-vanishing $v_\phi$)
\eq{N_{,\phi}(N)=N_{,\phi}(N_0)+e^{\int_{N_0}^Nu_\phi(N)\dd N}\int_{N_0}^Nv_\phi(N)e^{-\int_{N_0}^Nu_\phi(N)\dd N}\dd N.\label{eq-Calc-Nphi0}}
Here $N_{,\phi}(N_0)$ is the value of $N_{,\phi}$ at $N=N_0$, which serves as the boundary condition~\footnote{Although in \refeq{Calc-Nphi0} the integral of $u_\phi$ may give readers the illusion that it can be worked out analytically for any potential, it actually \emph{cannot} because $u_\phi$ is a function of both $\phi$ and $\psi$, neither of which remains constant. Therefore one cannot simply change the integral variable to $\phi$ or $\psi$  only.}.

\ssecs{Second order perturbations}
In certain cases, such as in the case of  separable potentials~\cite{GarciaBellido:1995qq,Vernizzi:2006ve}, the integrals in \refeq{Calc-Nphi0} can be worked out analytically, so $N_{,\phi}$ becomes a simple function of $\phi$ and $\psi$. For such cases the second order derivatives $N_{,\phi\phi}$, $N_{,\phi\psi}$ and $N_{,\psi\psi}$ can be derived by simply differentiating $N_{,\phi}$ and $N_{,\psi}$, while keeping aware of nonzero $N_{0,\phi}$ for general end of 
inflation and/or  post-inflationary conditions, such as in the case of reheating or preheating. In general, such integrals cannot be solved analytically. To calculate the second order derivatives for any potential, here we provide a general framework.

By definition, we decompose the second derivative along $\phi$ direction to the curvature and the isocurvature directions
\eq{N_{,\phi\phi}\equiv\frac{\partial N_{,\phi}}{\partial\phi}=N_{,\phi}'N_{,\phi}+\dot N_{,\phi} C_{,\phi}.\label{eq-Seco-Nphiphi}}
Since we already know the first term on the r.h.s, we only need to calculate the second term. To solve $\dot N_{,\phi}$ we can take dots on \refeq{Calc-Nphip} and utilize the interchangability between primes and dots. We then obtain the equation for $\dot N_{,\phi}$,
\eq{\dot N_{,\phi}'=u_\phi\dot N_{,\phi}+(u_{\phi,\phi}N_{,\phi}+v_{\phi,\phi})f_\phi+(u_{\phi,\psi}N_{,\phi}+v_{\phi,\psi})f_\psi.\label{eq-Seco-Nphip}}
Note that {\it dot} denotes derivative w.r.t $C$. 
This equation holds the same form as that of \refeq{Calc-Nphip}, and the solution should have the same integrated form as that of \refeq{Calc-Nphi0}. In \refeq{Seco-Nphiphi} and \refeq{Seco-Nphip}, $C_{,\phi}$ and $f_\phi$ can be solved by simple integration through
\eqa{
C_{,\phi}'&=&u_\phi C_{,\phi},\label{eq-Seco-Cphi}\\
f_\phi'&=&\left(\etaa-2\epa^2-\frac{N_{,\phi}}{N_{,\psi}}(\etab-2\epa\epb)\right)f_\phi.\label{eq-Seco-fphi}}

We can also obtain $\dot N_{,\phi}$ by differentiating \refeq{Calc-Nphi0} w.r.t $C$. It is expected to have exactly the same solution as that of in \refeq{Seco-Nphip}, as long we keep in mind that $N_0$ may also be a function of $C$ (if the boundary is not on a uniform-$N$ hypersurface), so $\dot N_0\neq 0$. Similarly, we can also calculate
\eq{N_{,\phi\psi}\equiv\frac{\partial N_{,\phi}}{\partial\psi}=N_{,\phi}'N_{,\psi}+\dot N_{,\phi} C_{,\psi}.}

Although we are capable of presenting the full integrated form for $N_{,\phi\phi}$, we are not expressing this in the current paper for obvious reasons --- the expressions are long and not practical for general non-separable potentials and we are not using the full integrated form in this paper. However, they do show that there exist analytical solutions for two-field slow roll inflation even for non-separable potentials albeit in the integrated form. In practice when calculating models with non-integrable $N_{,\phi}$ (in \refeq{Calc-Nphi0}), differential equations \refeq{Calc-Nphip}, \refeq{Seco-Nphip}, \refeq{Seco-Cphi}, and \refeq{Seco-fphi} are actually far more useful --- we can combine them with the background equations of motion and obtain all of them in a single run by numerically solving the differential equations \emph{backwards} from the end of inflation to the Hubble exit with an appropriate boundary condition.

Regarding the boundary conditions, the derivation for a general end of inflation condition is given in the appendix \refsec{Boundary conditions}. For separable potentials by sum or product, the integrals can be worked out analytically, and therefore the results can be simplified and extended. Such calculations are shown in the appendix \refsec{Separable potentials}.

\ssecs{Cosmological observables and non-Gaussianity}
We now have all the parameters required for deriving the cosmological observables. The power spectrum of curvature perturbations is equal to the power spectrum of $\delta N$ on large scales, which means
\eq{P_\zeta=P_{\delta N}\equiv N_{,\phi}^2P_{\delta\phi}+N_{,\psi}^2P_{\delta\psi}=\frac{H^2}{4\pi^2}(N_{,\phi}^2+N_{,\psi}^2).\label{eq-CObs-P}}
The scalar spectral index, $n_s$, is defined as
\eq{n_s-1\equiv-\frac{\dd\ln P_\zeta^2}{\dd N}=-2\epsilon^2-\frac{2(N_{,\phi}'N_{,\phi}+N_{,\psi}'N_{,\psi})}{N_{,\phi}^2+N_{,\psi}^2},\label{eq-CObs-ns}}
whose running can be determined by simple differentiation while using \refeq{Calc-Nphip}. The amplitude of bi-spectrum is characterized by the parameter $\fNL$, see~\cite{Lyth:2005fi, Seery:2005gb}:
\eqa{
\fNL &\approx &\frac{5}{6}\frac{\sum_{i,j}N_iN_jN_{ij}}{\left(\sum_iN_i^2\right)^2}\nonumber\\
&=&\frac{5}{6}\,\frac{N_{,\phi}^2N_{,\phi\phi}+N_{,\psi}^2N_{,\psi\psi}+2N_{,\phi}N_{,\psi}N_{,\phi\psi}}{(N_{,\phi}^2+N_{,\psi}^2)^2}.\label{eq-CObs-fNL}}
For tri-spectrum, we can only calculate up to second order in this formalism, which gives one of the two tri-spectrum parameters, see \cite{Byrnes:2006vq}
\eq{\tNL\equiv\frac{\sum_{i,j,k}N_iN_jN_{ik}N_{jk}}{\left(\sum_iN_i^2\right)^3}.}
We define the non-adiabaticity parameter $\widetilde\alpha$ for field perturbations at the end of inflation,
\eq{\frac{\widetilde\alpha}{1-\widetilde\alpha}\equiv\left.\frac{P_{\widetilde S}}{P_{\widetilde\zeta}}\right|_e,}
where $P_{\widetilde S}$ and $P_{\widetilde\zeta}$ indicate the power spectra of field perturbations in entropy and curvature directions respectively, and the subscript  $e$ denotes the end of inflation. The correlation parameter is defined from the cross-correlation power spectrum $P_{\widetilde S\widetilde\zeta}$, as
\eq{\widetilde\beta\equiv-\left.\frac{P_{\widetilde S\widetilde\zeta}}{\sqrt{P_{\widetilde S}P_{\widetilde\zeta}}}\right|_e.}
We calculate the field perturbations along the entropy and the curvature directions at the end of inflation from those at the Hubble exit through $\delta N$ and $\delta C$, which remain constant during inflation. This yields
\eqa{
\frac{\widetilde\alpha}{1-\widetilde\alpha}&=&\left.\frac{4\pi}{\mpl^2\epsilon^2(C_{,\phi}^2+C_{,\psi}^2)}\right|_e\frac{C_{,\phi}^2+C_{,\psi}^2}{N_{,\phi}^2+N_{,\psi}^2},\label{eq-CObs-alphat}\\
\widetilde\beta&=&\pm\frac{N_{,\phi}C_{,\phi}+N_{,\psi}C_{,\psi}}{\sqrt{(N_{,\phi}^2+N_{,\psi}^2)(C_{,\phi}^2+C_{,\psi}^2)}},\label{eq-CObs-betat}}
where $\pm$ should be chosen as the sign of $(\epa f_\psi-\epb f_\phi)|_e$. The definition of $\widetilde\alpha$ and $\widetilde\beta$ are in the field space, unlike $\alpha$ and $\beta$ defined as power spectrum ratios of energy density perturbations in Ref.~\cite{Komatsu:2008hk}. This is because one cannot split the total energy density to field energy densities if they are coupled. However, for separable potentials by sum in the form \refeq{Sep-Sum}, fields are uncoupled so we can calculate the power spectra of non-adiabaticity perturbation and its correlation with curvature perturbation in energy density. See  the derivation in the appendix \refssec{Separable by sum}.

We also define the power spectrum ratio between adiabatic and curvature perturbations:
\eq{ 0\leqslant\gamma\equiv\frac{(\epa N_{,\phi}+\epb N_{,\psi})^2}{\epsilon^2(N_{,\phi}^2+N_{,\psi}^2)}\leqslant 1,\label{eq-CObs-gamma}}
where $\gamma\rightarrow 0$ means the whole curvature perturbation comes from the conversion of the entropy perturbation and vice versa. Strictly speaking the approximation of taking the adiabatic perturbation as the curvature perturbation (i.e.\ the so-called \emph{adiabatic limit}) is only allowed when $\gamma\rightarrow1$ is satisfied both at the time of Hubble exit and onwards.

\secs{Two-field inflationary models}
We will demonstrate our formalism through a simple model of two-field inflation in which two scalar fields $\phi$ and $\psi$ prevent the waterfall field $\chi$ from falling~\footnote{Similar studies were performed in Refs.~\cite{Burgess:2005sb,Choi:2012he,Byrnes:2008zy,Huang:2009vk,Sasaki:2008uc,Naruko:2008sq}. Note that in this paper $\phi$ and $\psi$ are not charged under the Standard Model gauge group or its minimal extensions~\cite{Allahverdi:2006iq,Allahverdi:2006cx,Allahverdi:2006we}. Here we treat them as singlets, and we also assume that $\phi$, $\psi$ and $\chi$ all decay into the Standard Model radiation without making any justification.  In this respect this model serves at best as a simple {\it toy} model which can {\it only} mimic the success of CMB perturbations.}. The potential  is given by:
\eq{V=V_0\left(1-\frac{\chi^2}{v^2}\right)^2+(g_\phi^2\phi^2+g_\psi^2\psi^2)\chi^2+\widetilde V(\phi,\psi).\label{eq-Eg-V}}
Inflation is dominated by the constant potential energy density, $V_0$, with $\chi$ sitting at \vev $0$ during inflation. Both $\phi$ and $\psi$ are slow rolling initially~\footnote{We do not study the initial condition problem for this model, we expect that the initial condition problem for this model will be similar to that of any hybrid model of inflation, see~\cite{Clesse:2009ur}.}. Inflation ends once the effective mass of $\chi$ becomes negative, which gives the end of inflation condition as
\eq{g_\phi^2\phi_e^2+g_\psi^2\psi_e^2=2V_0/v^2,}
where the subscript $e$ means the end of inflation. 

\fig{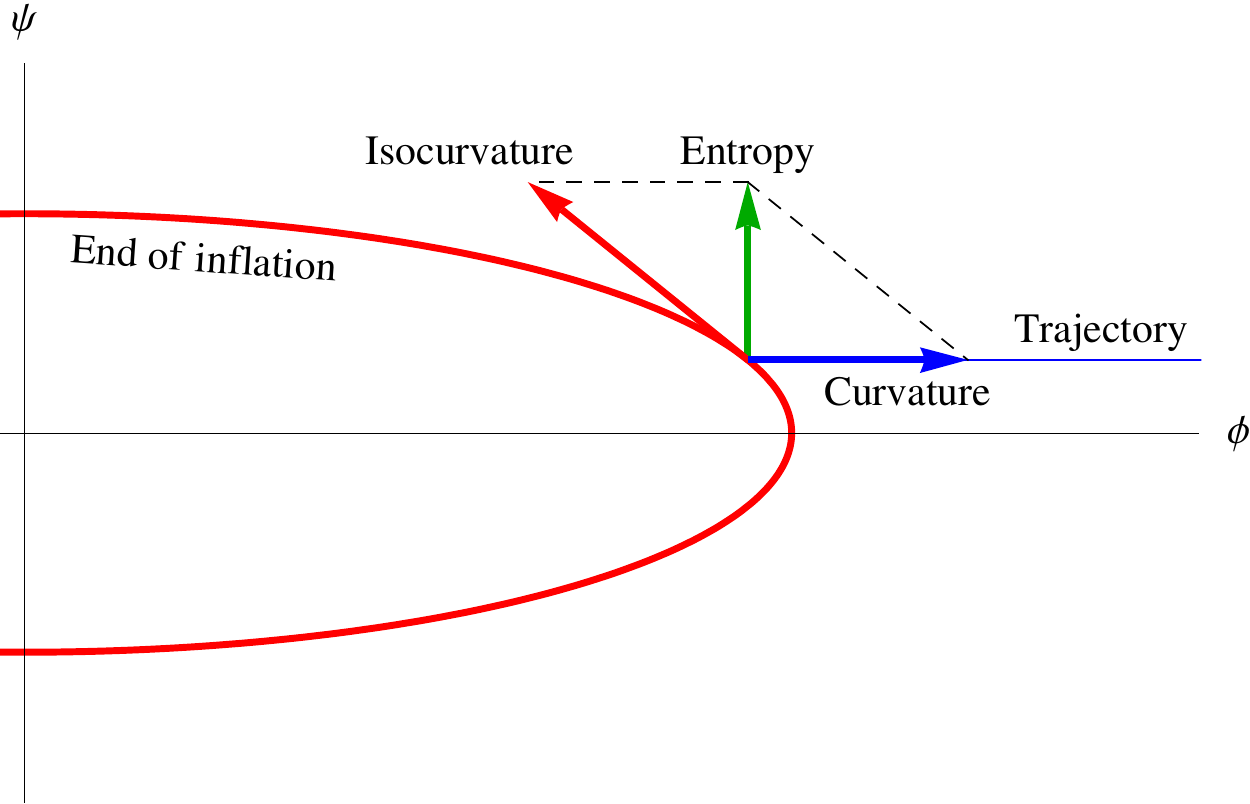}{A schematic figure shows how the entropy perturbations are converted into the curvature (and isocurvature) perturbations at the end of inflation, by assuming that no significant conversion happens after inflation. For $\delta\psi>0$, the trajectory is shifted upwards slightly, leading to a longer inflation and therefore $\delta N>0$. A large non-Gaussianity may be generated if the boundary is very curved at its intersection with the trajectory.}{\defaultfigsize}{}

The end of inflation happens by waterfall, so the trajectory of inflation and the end of inflation boundary can form a sharp angle, which allows a significant transfer from the entropy to curvature perturbations at the end of inflation. Large non-Gaussianity can also be achieved through this mechanism when the couplings $g_\phi$ and $g_\psi$ differ significantly. This also requires that the entropy perturbation to remain non-vanishing during the last 60 e-foldings of exponential expansion, which means one field has to be lighter than the other. This is shown in \refig{Eg-Traj}. 

Since during inflation the effective potential is just $V=V_0+\widetilde V(\phi,\psi)$, whether the potential is separable depends on $\widetilde V$. In the following, we will first demonstrate a separable case, giving both analytical and numerical results for comparison, and then provide a non-separable potential model.

\subsection{Separable potentials}
For the separable potential case, we take the double quadratic potential of the form:
\eq{\widetilde V(\phi,\psi)=m_\phi^2\phi^2+m_\psi^2\psi^2.\label{eq-Eg-Vt}}
We then use the expression of $N_{,\phi}$ for separable potential case by substituting \refeq{Eg-Vt} into \refeq{Sep-Sump}. We also take the approximation that $\psi$ is much lighter than $\phi$, so it moves very slowly during the last 60 e-foldings of inflation. The dominating terms, at first order are:
\eqa{N_{,\phi}&=&\frac{4\pi V_0}{\mpl^2m_\phi^2\phi}\label{eq-Eg-aNa},\\
N_{,\psi}&=&\frac{4\pi V_0}{\mpl^2m_\phi^2\psi}\tan^2\theta,\label{eq-Eg-aNb}}
where $\theta$ characterizes the end point of inflation on the elliptical boundary, and it is defined as:
\eq{\tan\theta\equiv\frac{g_\psi\psi_e}{g_\phi\phi_e}.\label{eq-Eg-theta}}
For the case $m_\phi^2/m_\psi^2\gg\tan^2\theta$ which we are interested in, the differential relations hold:
\eq{\frac{\partial\theta}{\partial\phi}=-\frac{m_\psi^2}{m_\phi^2}\frac{\tan\theta}{\phi},\hspace{0.8in}
\frac{\partial\theta}{\partial\psi}=\frac{\tan\theta}{\psi}.}
The second order derivatives can then be calculated through differentiating \refeq{Eg-aNa} and \refeq{Eg-aNb}, giving the dominant term
\eq{N_{,\psi\psi}=\frac{4\pi V_0(2\tan^2\theta+1)\tan^2\theta}{\mpl^2m_\phi^2\psi^2}.\label{eq-Eg-aNbb}}
The cosmological observables can then be derived from substituting \refeq{Eg-aNa}, \refeq{Eg-aNb} and \refeq{Eg-aNbb} into \refeq{CObs-P}, \refeq{CObs-fNL} and \refeq{CObs-gamma}, giving
\eqa{
P_\zeta&=&\frac{128\pi^3V_0^3}{3\mpl^6m_\phi^4}\left(\frac{1}{\phi^2}+\frac{1}{\psi^2}\tan^4\theta\right),\\
\fNL&=&\frac{5\mpl^2m_\phi^2(2\tan^2\theta+1)\tan^6\theta}{24\pi V_0(\tan^4\theta+\psi^2/\phi^2)^2},\\
\gamma&=&\frac{(m_\phi^2+m_\psi^2\tan^2\theta)^2}{m_\phi^4+m_\psi^4\tan^4\theta+m_\phi^4\tan^4(\theta)\phi^2/\psi^2+m_\psi^4\psi^2/\phi^2}.\label{eq-Eg-gamma}}
The slow roll approximation gives the evolution of fields
\eq{\phi=\phi_ee^{\frac{m_\phi^2\mpl^2}{4\pi V_0}N},\hspace{0.8in}\psi=\psi_ee^{\frac{m_\psi^2\mpl^2}{4\pi V_0}N}.}

Since $m_\psi\ll m_\phi$, during inflation $\psi$ remains almost constant and $\phi$ drops gradually, while all other parameters remain constant. Therefore as $N$ gradually decreases, we obtain an increasing trend of $P_\zeta$ (which remains within $2\sigma$ of the current bound on the amplitude of the power spectrum~\cite{Komatsu:2010fb}), and a decreasing $\fNL$, while a slowly increasing $n_s-1$, and  $\gamma$ depending on the parameter choice (which is increasing here).

\figa{
\begin{tabular}{c@{\hspace{0.08\textwidth}}c}
\figi{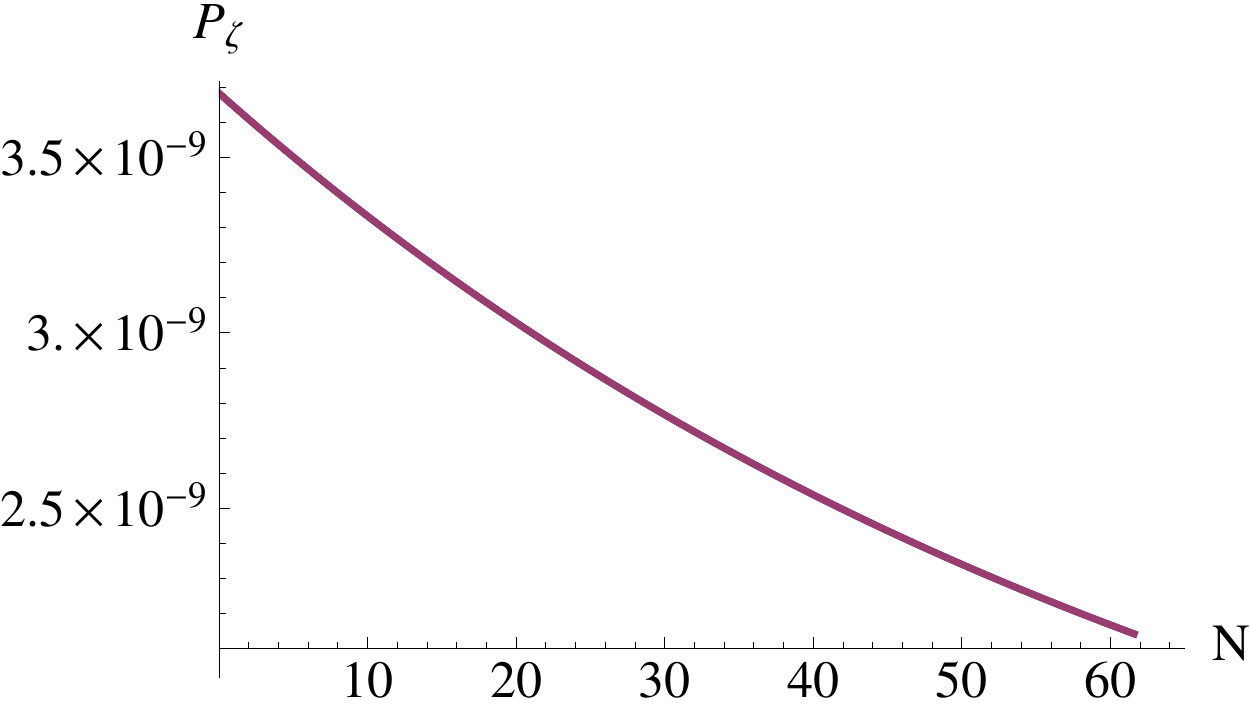}{Scalar power spectrum}{0.45}&\figi{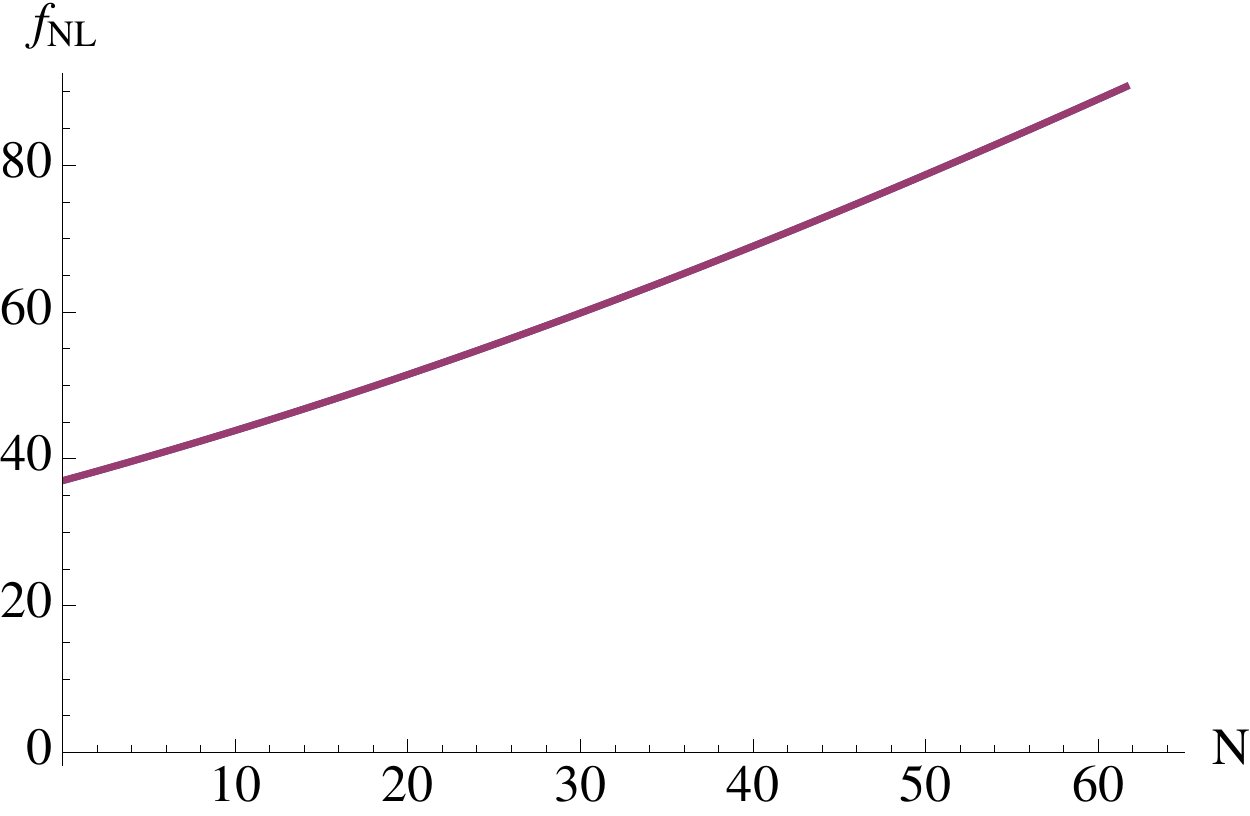}{Non-Gaussianity}{0.45}\\
\figi{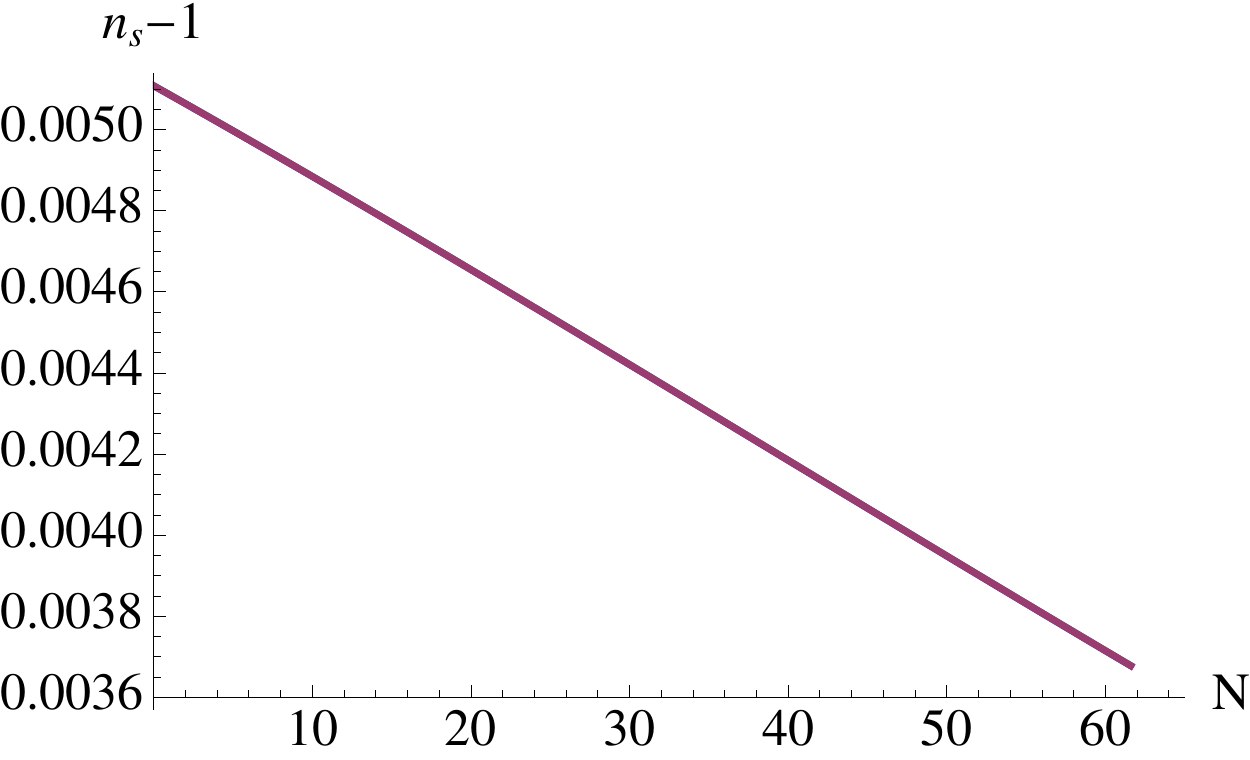}{Scalar spectral tilt}{0.45}&\figi{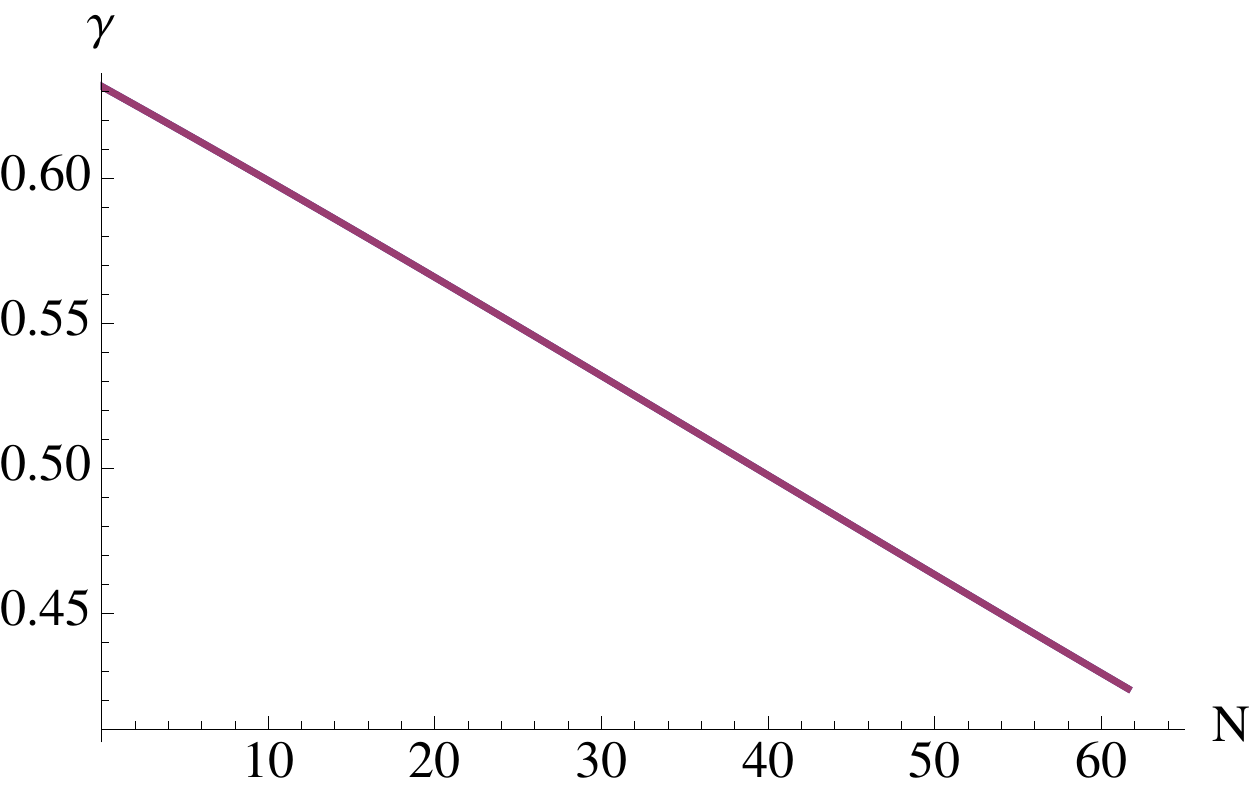}{Adiabatic-curvature ratio}{0.45}
\end{tabular}
}{Red and blue curves are analytical and numerical results respectively. They are too close to be distinguished from each other, showing the analytical approximations as compared to the numerical results for the physical parameters~\refeq{parameter-fix}.\label{fig-NMulti}}{}
Our formalism allows us to numerically solve the differential equations \refeq{Calc-Nphip}, \refeq{Seco-Nphip}, \refeq{Seco-Cphi}, and \refeq{Seco-fphi}. Unlike the finite-difference method applied in  Ref.~\cite{Vernizzi:2006ve}, this method is not plagued by the dilemma of balancing between error sources~\footnote{In the finite difference method errors come from the inaccuracies from higher order contributions and imprecisions from small relative differences, whose strengths are inversely correlated. If the pivot scale exits the Hubble patch during inflation leaving $\fNL\sim100$ and $\phi N_{,\phi}=10^xN\subs{pivot}$, where $N\subs{pivot}\sim 50$ has 10 significant figures from solving the equations numerically, then a rough estimation of the error for uncorrected finite difference method gives the number of significant figures of $N_{,\phi\phi}$ at most $1.75-|x|/4$.}, so it typically has much more precise results at the second order and the errors are easily controllable.

For this model, numerical results are shown for the parameters in \refeq{Eg-V} and \refeq{Eg-Vt}:
\eq{
\begin{array}{l@{\hspace{0.4in}}l@{\hspace{0.4in}}l}
m_\phi=3.03\times10^{10}\gev,&g_\phi=1.13\times10^{-3},&V_0=(1.09\times10^{15}\gev)^4,\\
m_\psi=9.57\times10^9\gev,&g_\psi=1.79\times10^{-1},&v=1.22\times10^{16}\gev. \label{eq-parameter-fix}
\end{array}}
For these parameters if the fields start from $\psi\ll\phi$, we can obtain a large non-Gaussianity. As an example,  here we consider  $\phi=1.95\times10^{17}\gev$ and $\psi=3.91\times10^{12}\gev$, which result in the total scalar power spectrum, scalar spectral tilt, non-Gaussianity and the adiabatic-curvature ratio as functions of remaining e-foldings shown and compared with the analytical solutions in \refig{NMulti}. In this case, we also obtain the largest $\tNL\sim20000$.

\figa{
\begin{tabular}{c@{\hspace{0.08\textwidth}}c}
\figi{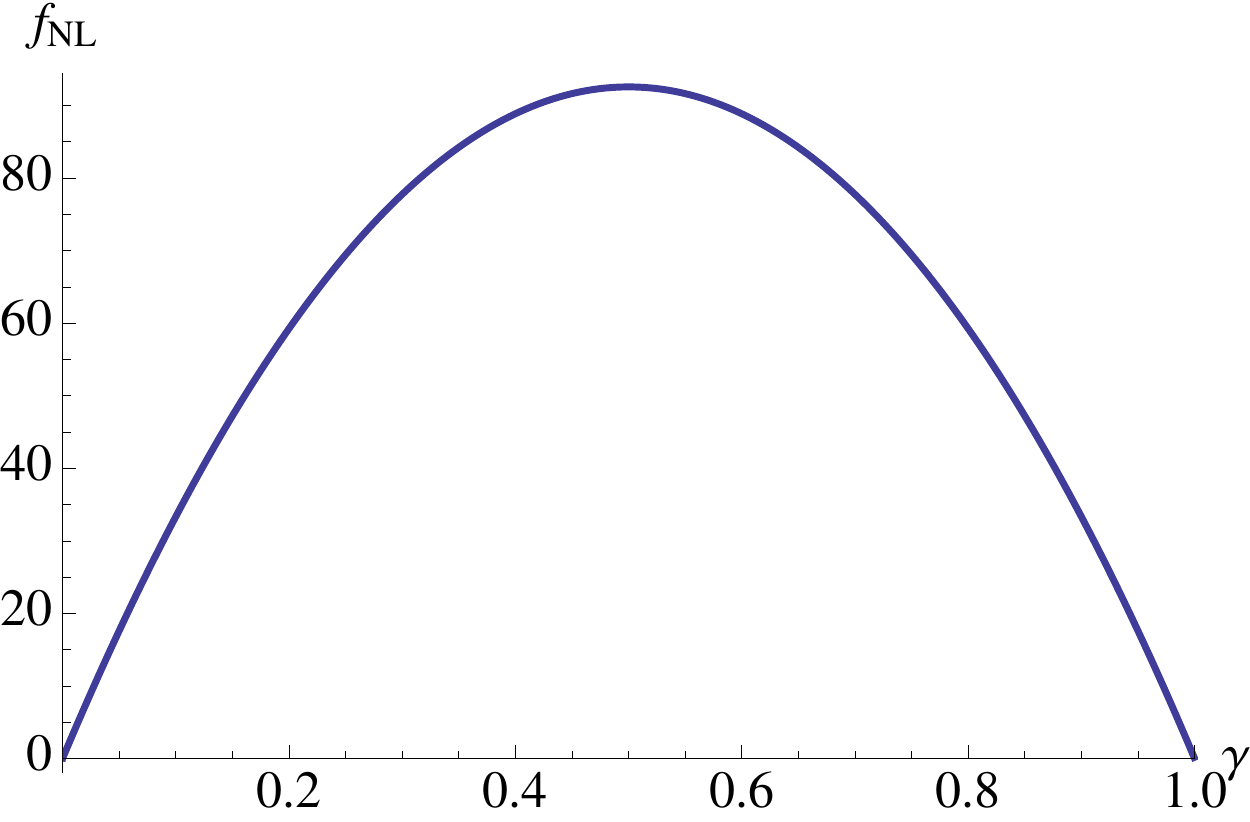}{Bi-spectrum parameter}{0.45}&\figi{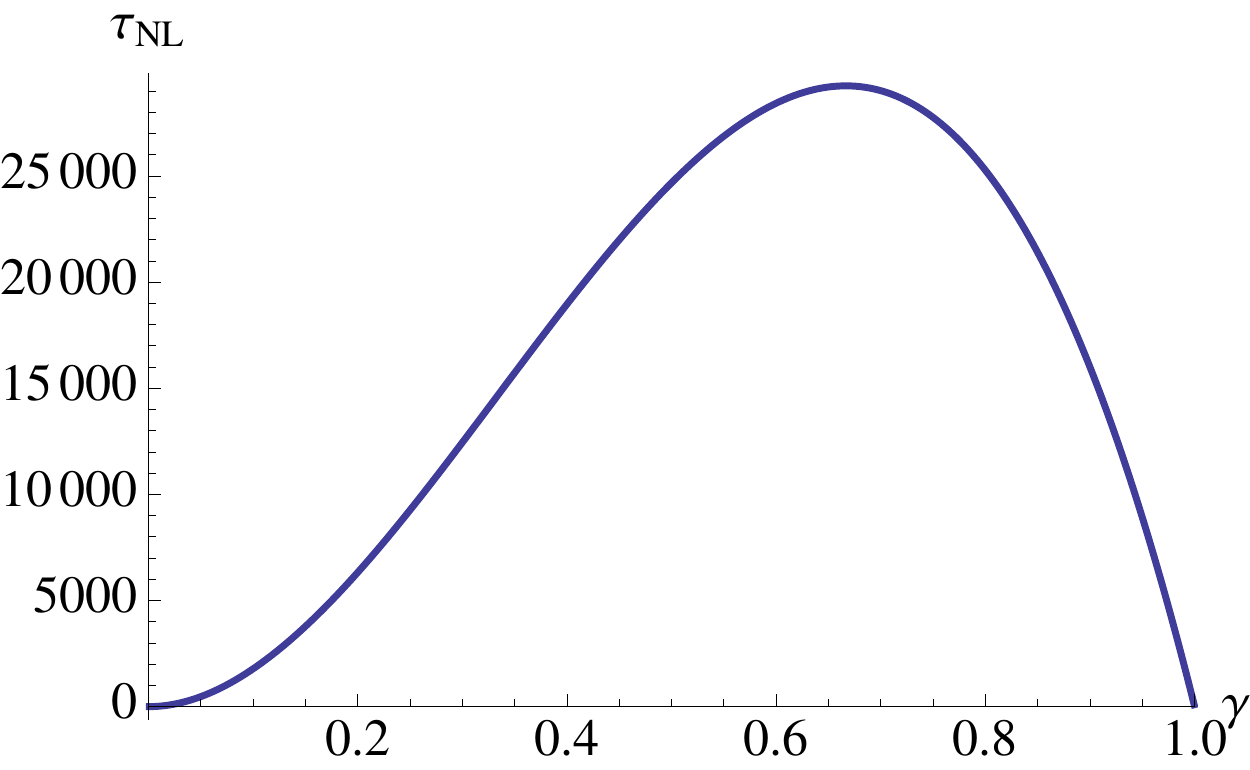}{Tri-spectrum parameter}{0.45}
\end{tabular}
}{Numerical results of bi- and tri-spectrum parameters with fixed model parameters in \refeq{parameter-fix}, but varying initial conditions of $\phi$ and $\psi$. See \refeq{Eg-theta} and \refeq{Eg-gamma}. \label{fig-NMultig}}{}
It is worth mentioning here that if we change the initial condition while fixing all the parameters, we are able to get $\fNL$ and $\tNL$ as functions of $\gamma$ at the pivot scale, corresponding to $k=0.002\mathrm{\ Mpc}^{-1}$, which are shown in \refig{NMultig}. We can see that under these parameters an interesting result is that we obtain the largest $\fNL$ when $\gamma=0.5$, i.e. when adiabatic and entropy perturbations give equal contributions to the power spectrum of the curvature perturbations.

This effect is easily understood qualitatively. A dominant contribution to $\fNL$ comes from the entropy perturbations so we expect $\fNL$ to grow as $\gamma$ decreases from $1$. As long as changing $\gamma$ does not significantly affect the value of $\phi$ at the Hubble exit of the relevant scale, the relation $N_\phi^2+N_\psi^2\propto\gamma^{-1}$ holds which contributes a $\gamma^2$ coefficient to $\fNL$. Also, one can find the same conclusion by solving $\theta$ as a function of $\gamma$ and substituting it into $\fNL$, which then precisely leads to $\fNL\propto\gamma(1-\gamma)$. For $\tNL$, a similar relation holds $\tNL\propto\gamma^2(1-\gamma)$.

\ssecs{Non-separable potentials}
As a non-separable example, we consider a simple Logarithmic potential
\eq{\widetilde V(\phi,\psi)=V_1\ln\left(\frac{\phi^2+\lambda^2\psi^2}{\mpl^2}\right),}
where $V_1$ is the energy scale of the two fields, and $\lambda$ is the parameter which characterizes the mass ratio between them.

\figa{
\begin{tabular}{c@{\hspace{0.08\textwidth}}c}
\figi{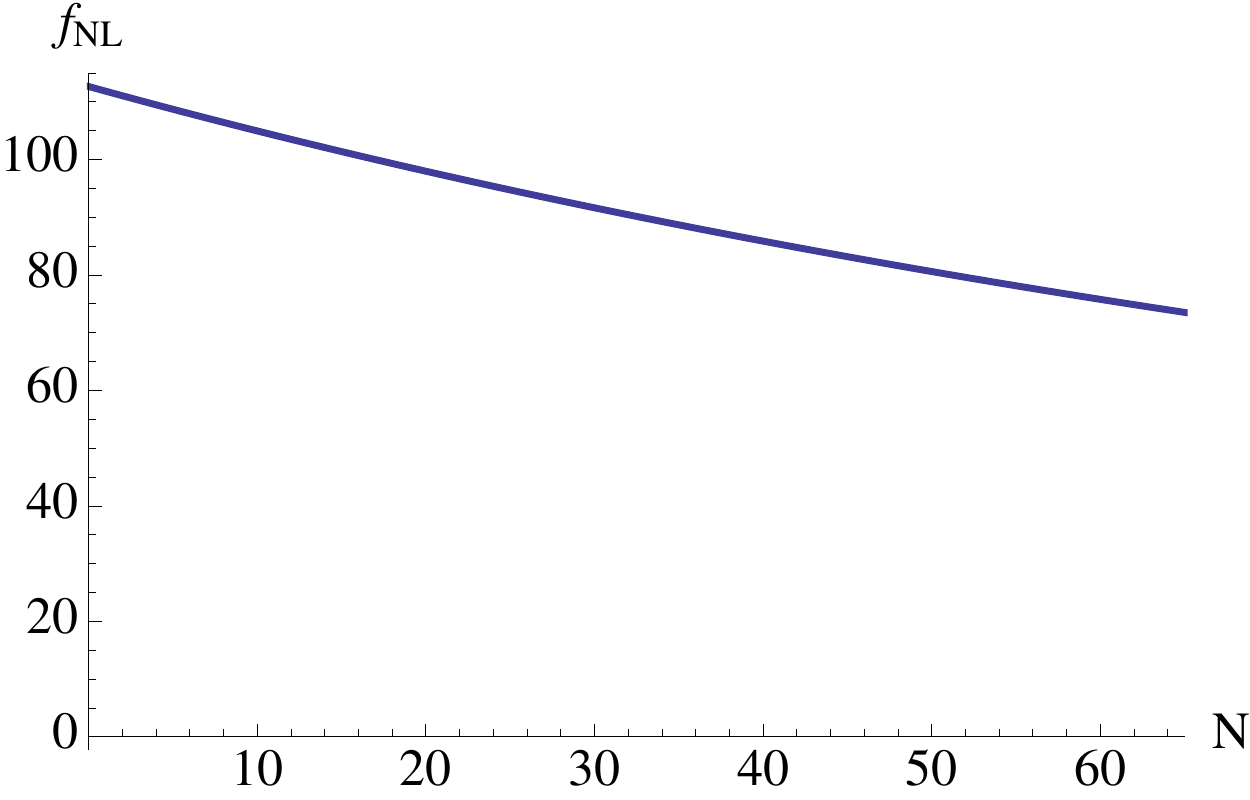}{Non-Gaussianity}{0.45}&\figi{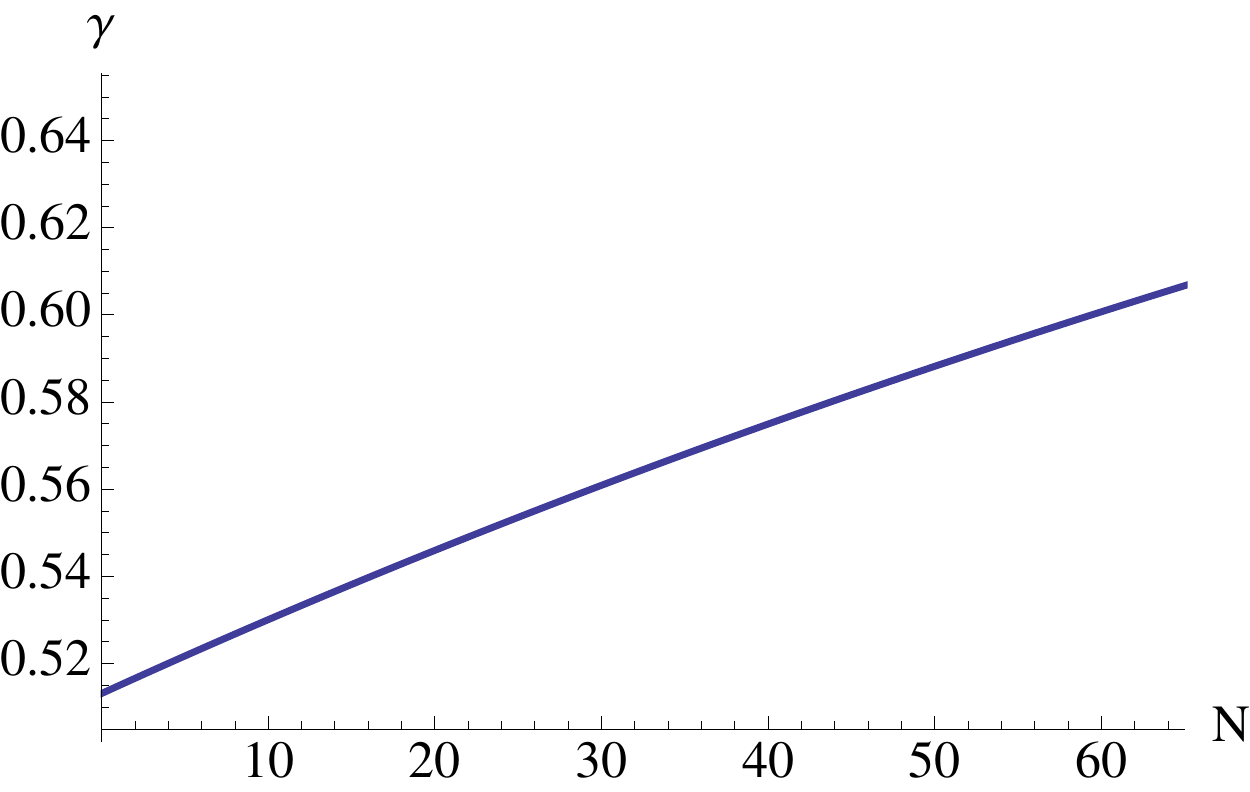}{Adiabatic-curvature ratio}{0.45}\\
\figi{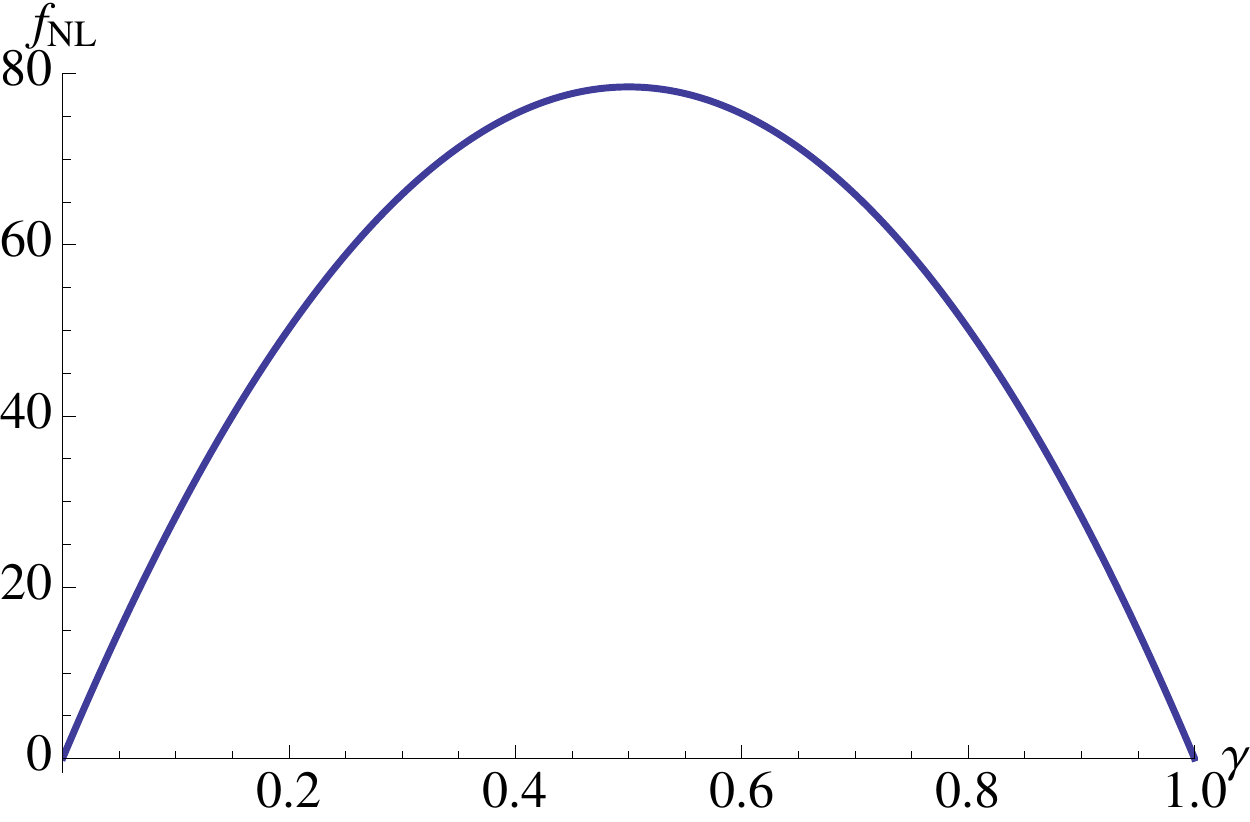}{Bi-spectrum}{0.45}&\figi{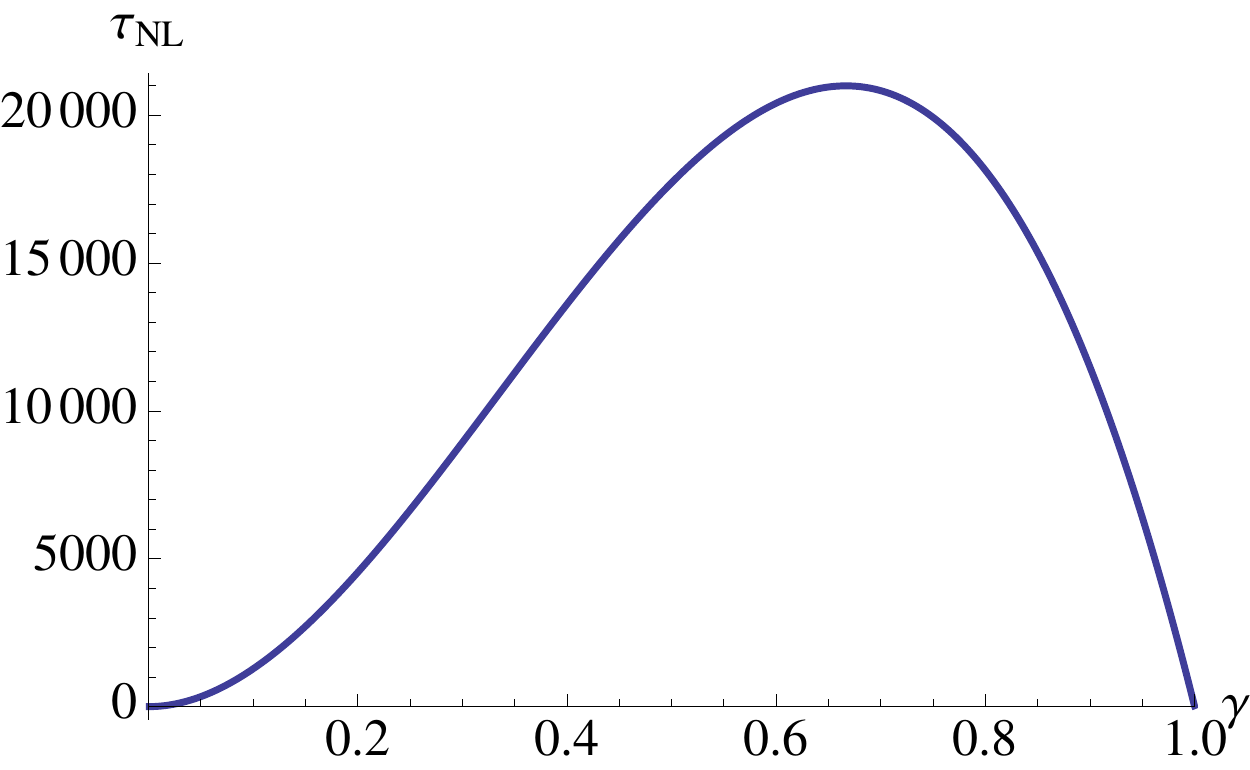}{Tri-spectrum}{0.45}
\end{tabular}
}{Numerical results for a  non-separable potential is shown above. For \refig{Eg2-fNL} and \refig{Eg2-Gamma}, we use the fixed initial conditions. For \refig{Eg2-fNLg} and \refig{Eg2-tNLg}, we allow the initial conditions of $\phi$ and $\psi$ to vary while fixing the e-folding we are interested in. The physical parameters are given by Eq.~(\ref{parameter-fix-1}). \label{fig-NMulti2}}{}
For the parameters shown below, we obtain the right amplitude and the tilt in the power spectrum, but also 
large non-Gaussianity
\eq{
\begin{array}{l@{\hspace{0.5in}}l@{\hspace{0.5in}}l}
\phi=1.59\times10^{16}\gev,&g_\phi=3.16\times10^{-5},&V_0=(1.83\times10^{14}\gev)^4,\\
\psi=7.18\times10^{10}\gev,&g_\psi=1.30\times10^{-2},&V_1=(2.58\times10^{12}\gev)^4,\\
v=1.22\times10^{17}\gev,&\lambda^2=0.1.\label{parameter-fix-1}
\end{array}}
The numerical solutions of \refeq{Calc-Nphip}, \refeq{Seco-Nphip}, \refeq{Seco-Cphi}, and \refeq{Seco-fphi} are shown in \refig{NMulti2}, with $P_\zeta\approx2.4\times 10^{-9}$ and $n_s-1\approx-1.3\times10^{-3}$, which are quite similar to \refig{NMulti} and \refig{NMultig}. This is reasonable as they rely on the same end-of-inflation mechanism to generate large non-Gaussianity. From this example, it is evident that the numerical method also works well with non-separable potentials.

We note that the $\fNL$'s and $\tNL$'s have similar shapes in \refig{NMultig} and \refig{NMulti2}. The large non-Gaussianity arises due to the end of inflation boundary condition rather than the inseparability of the potential. Although here the inseparability is not playing an important role and we only use it to demonstrate the capability of the formalism, non-separable potentials still provide distinct features in many cases, such as multi-stream inflation\cite{Duplessis:2012nb} and the study of the effects of couplings for N-flation\cite{Dimopoulos:2005ac,Easther:2005zr}. In such models this formalism will be helpful in the prediction of cosmological observables.

\secs{Comparison with related works}
At first, Yokoyama et al. proposed to calculate the evolution of $N_{,\mu}$ and $N_{,\mu\nu}$ backwards from the end of inflation to the Hubble exit in Refs.~\cite{Yokoyama:2007uu,Yokoyama:2007dw}. They used the transfer matrix method, and the analytical representations are mostly formal. In particular, their derivation of $N_{,\mu\nu}$ has two levels of embedded integrals which are unfavored by numerical calculations. Recently Mulryne et al. managed to evolve the distribution of field perturbations after their Hubble exit, which is called \emph{moment transport equation} method~\cite{Mulryne:2010rp,Mulryne:2009kh}. These methods are physically equivalent but have different advantages --- in a single run the backward formalism is capable of calculating the amplitudes of all modes at a specific time, which is the case for CMB, while the moment transport equation method evolves a specific mode all through inflation,  deriving its amplitude at any time.

This paper has the same idea with those by Yokoyama et al, but attempts to understand the perturbations in a geometrical approach. Therefore the derivations are more visual and straightforward than Refs.~\cite{Yokoyama:2007uu,Yokoyama:2007dw}. In addition, two improvements have been made in this paper. As a result of the geometrical approach and the additional parameter $C$, we are able to address isocurvature perturbations within the framework of $\delta N$ formalism, which is otherwise considered difficult as mentioned in Ref.~\cite{Seery:2012vj}. We also use a generic end of inflation condition, which allows post-inflation and/or end-of-inflation mechanisms to generate significant curvature perturbations. For two-field inflation, we also find the equations for derivatives of $N$ (e.g.\ \refeq{Calc-Nphi0} and \refeq{Seco-Nphip}) can be separated, providing a higher efficiency in cases where we know in advance whose perturbation will dominate.

We would also like to briefly mention here one specific advantage of the backward formalism (e.g.\ this paper) over moment transport equation method. As shown in \refssec{Background and first order perturbations} and \refssec{Second order perturbations}, derivatives of $N$ and $C$ are independent and can be solved separately. This means we can pick out only the one(s) we are interested in and save time by discarding the rest, which however the moment transport equation method is unable to achieve. The lower time complexity will become obvious when we have many fields such as in N-flation.\footnote{For $n$-field calculation, Refs.\ \cite{Yokoyama:2007uu,Yokoyama:2007dw} introduced $\Theta^I$ to reduce the number of integrals to $O(n)$. Although the $n$-field extension of this formalism has $O(n^2)$ integrals, the total time complexity remains the same with that of \cite{Yokoyama:2007uu,Yokoyama:2007dw}. This is because each integral in \cite{Yokoyama:2007uu,Yokoyama:2007dw} needs $O(n^2)$ time to calculate every sum in each integral while (the extension of) this formalism only needs $O(n)$. Therefore both methods have a total $O(n^3)$ time complexity and Refs.\ \cite{Yokoyama:2007uu,Yokoyama:2007dw} don't have any advantage over this formalism in this sense.}

We have compared the analytical differential equations of $N$'s derivatives with those in Refs.~\cite{Yokoyama:2007uu,Yokoyama:2007dw,Mulryne:2010rp,Mulryne:2009kh,Seery:2012vj}, and they match in every detail. Verifications have also been made through comparisons between papers about separable potentials and our reduced result in \refsec{Separable potentials}. 

\secs{Summary}
We have proposed a simple framework based on $\delta N$ formalism which deals with two-field slow roll inflation models. The formalism gives analytical integrated solutions up to second order perturbations for general non-separable potentials. In the case of separable potentials, they are exactly integrable. This formalism can be easily adapted for numerical solutions to non-separable potentials. In the examples we demonstrated, our results match the expectations of \cite{Jokinen:2005by,Naruko:2008sq,Sasaki:2008uc} that in order to obtain large $\fNL$ and $\tNL$, one must generate entropy perturbations during inflation which then convert to curvature perturbations non-adiabatically.

The major difference with previous studies is that we decompose the perturbations into curvature and isocurvature rather than adiabatic and entropy. Therefore it also enables the calculation of isocurvature perturbations while basing on $\delta N$ formalism. Although this paper only discusses two-field slow roll inflation, our formalism can also be extended to arbitrary number of fields with non-canonical kinetic terms, and also fast roll fields.

\acknowledgments{We would like to thank Ki-Young Choi, Kari Enqvist, David Mulryne, Misao Sasaki, David Seery, Philip Stephens and David Wands for helpful discussions. The work of AM is supported by STFC grant ST/J000418/1.}

\appendix

\secs{Boundary conditions}
Let us suppose that the two-field slow roll inflation stage ends at some general classical boundary, given by: $\sigma(\phi,\psi)=0$, which could be either due to violation of slow roll conditions, such as $\sigma=\epsilon^2-1$, or via some sudden phase transitions. Since the boundary need not necessarily be on a uniform-$N$(/isocurvature) hypersurface, we will assume \emph{on the boundary} the remaining e-foldings $N_0(\phi,\psi)$ is well-known, which has $\dd N_0/\dd C\ne 0$.

\fig{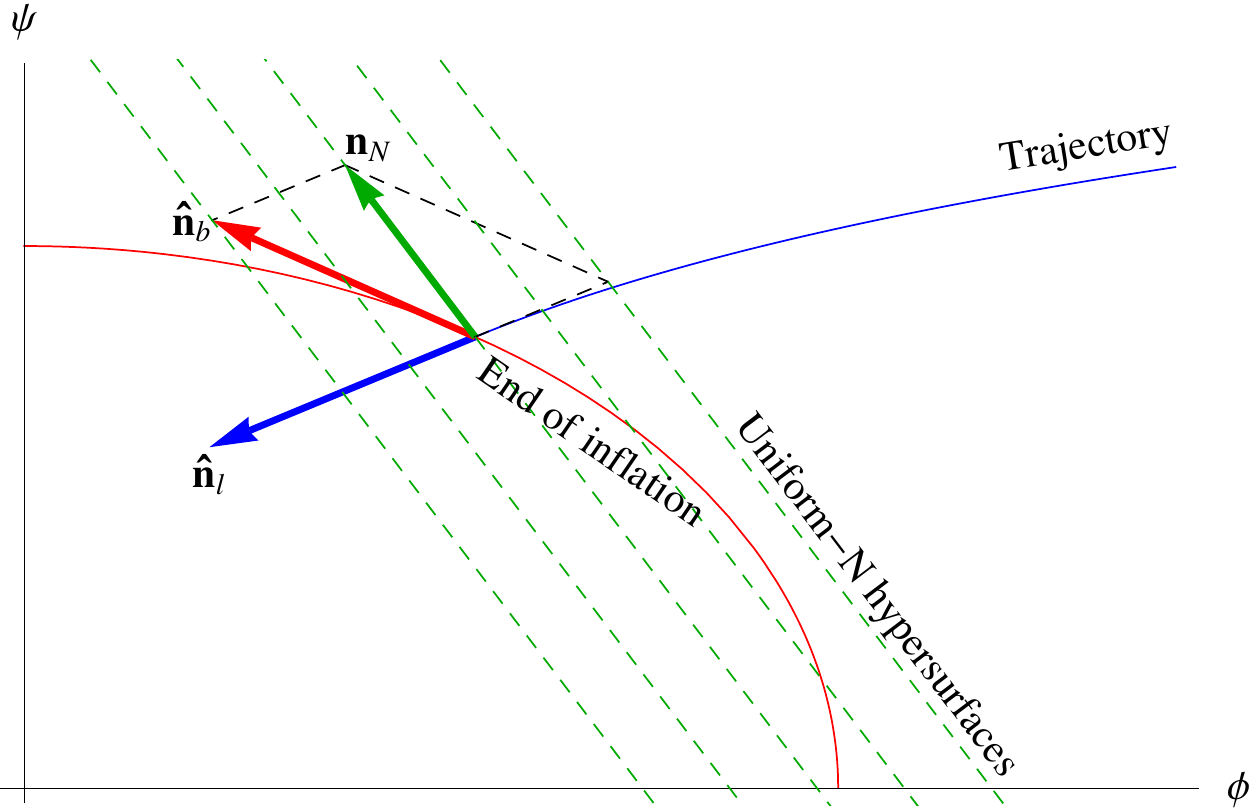}{A schematic figure which illustrates how $C$ is defined through uniform-$N$ hypersurfaces at the end of inflation. The blue and red curves are inflationary trajectory and the end of inflation boundary, respectively. The arrows in blue and red are unit vectors $\hatt{n}_l$ and $\hatt{n}_b$,  from which the green arrow $\mathbf{n}_N$, the vector in the uniform-$N$(/isocurvature) direction, can be constructed. The green dashed lines show the uniform-$N$ hypersurfaces.}{\defaultfigsize}{}
From \refig{Boundary}, we can work out the first and second order boundary conditions through vectors. Pick a point $(\phi,\psi)$ on the boundary and set $C=0$ here, the unit vector along the trajectory is then given by
\eq{\hatt{n}_l\equiv\frac{(\phi',\psi')}{\sqrt{\phi'{}^2+\psi'{}^2}}=\left(\frac{\epa}{\epsilon},\frac{\epb}{\epsilon}\right).}
Along the boundary, we have $\delta\sigma=\sigma_{,\phi}\delta\phi+\sigma_{,\psi}\delta\psi=0$, so its unit vector is
\eq{\hatt{n}_b\equiv\frac{(\sigma_{,\psi},-\sigma_{,\phi})}{\sqrt{\sigma_{,\phi}^2+\sigma_{,\psi}^2}}.}
To find a vector on the uniform-$N$ hypersurface, we construct an infinitesimal general vector
\eq{\mathbf{x}\equiv x_l\hatt{n}_l+x_b\hatt{n}_b.}
This displacement in the phase space then generates
\eq{\delta N=(N_{0,\phi},N_{0,\psi})\cdot\hatt{n}_bx_b+\frac{2\sqrt\pi}{\mpl\epsilon}x_l,\label{eq-Boun-deltaN}}
in which the first term on the r.h.s is the contribution outside(/after crossing) the boundary, and the second term comes within the boundary (during two-field slow roll inflation). 

By vanishing $\delta N$ in \refeq{Boun-deltaN}, we are able to find the vector on the uniform-$N$ hypersurface
\eq{\mathbf{n}_N=\hatt{n}_b-\frac{\epsilon\mpl(\sigma_{,\psi}N_{0,\phi}-\sigma_{,\phi}N_{0,\psi})}{2\sqrt{\pi(\sigma_{,\phi}^2+\sigma_{,\psi}^2)}}\hatt{n}_l.}
For a general displacement at the end of inflation, $\mathbf{x}\equiv(\delta\phi,\delta\psi)$, we \emph{define} $C$ (or $\delta C$) by decomposing it as
\eq{\mathbf{x}=\delta C\,\mathbf{n}_N+\widetilde x_l\hatt{n}_l|_{N=N_0}.\label{eq-Boun-Cdef}}
Then obviously the condition $\dd N/\dd C=\dd C/\dd N=0$ is automatically satisfied. From \refeq{Boun-Cdef}, we find the following boundary conditions by vanishing $\widetilde x_l$, $\delta\psi$, or $\delta\phi$
\eqa{
f_\phi(N_0)&\equiv&\dot\phi(N_0)=\frac{2\sqrt\pi\sigma_{,\psi}-(N_{0,\phi}\sigma_{,\psi}-N_{0,\psi}\sigma_{,\phi})\mpl\epa}{2\sqrt{\pi(\sigma_{,\phi}^2+\sigma_{,\psi}^2)}},\\
f_\psi(N_0)&\equiv&\dot\psi(N_0)=-\frac{2\sqrt\pi\sigma_{,\phi}+(N_{0,\phi}\sigma_{,\psi}-N_{0,\psi}\sigma_{,\phi})\mpl\epb}{2\sqrt{\pi(\sigma_{,\phi}^2+\sigma_{,\psi}^2)}},\\
N_{,\phi}(N_0)&=&\frac{2\sqrt\pi\sigma_{,\phi}/\mpl+(N_{0,\phi}\sigma_{,\psi}-N_{0,\psi}\sigma_{,\phi})\epb}{\epa\sigma_{,\phi}+\epb\sigma_{,\psi}},\\
C_{,\phi}(N_0)&=&\frac{\epb\sqrt{\sigma_{,\phi}^2+\sigma_{,\psi}^2}}{\epa\sigma_{,\phi}+\epb\sigma_{,\psi}},\\
N_{,\psi}(N_0)&=&\frac{2\sqrt\pi\sigma_{,\psi}/\mpl-(N_{0,\phi}\sigma_{,\psi}-N_{0,\psi}\sigma_{,\phi})\epa}{\epa\sigma_{,\phi}+\epb\sigma_{,\psi}},\\
C_{,\psi}(N_0)&=&-\frac{\epa\sqrt{\sigma_{,\phi}^2+\sigma_{,\psi}^2}}{\epa\sigma_{,\phi}+\epb\sigma_{,\psi}}.}
Boundary conditions for the second order perturbations can be derived from
\eq{\dot N_{,\phi}(N_0)=f_\phi(N_0)N_{,\phi\phi}(N_0)+f_\psi(N_0)N_{,\phi\psi}(N_0).}

\secs{Separable potentials}
\ssecs{Separable by sum}
Using the potential,
\eq{V(\phi,\psi)=U(\phi)+W(\psi),\label{eq-Sep-Sum}}
we will find the integral in \refeq{Calc-Nphi0} can now be worked out analytically
\eq{N_{,\phi}=\frac{U'(N_0)}{U'}N_{,\phi}(N_0)+\frac{8\pi}{\mpl^2U'}\bigl(U-U(N_0)\bigr).\label{eq-Sep-Sump}}
\Refeq{Seco-Cphi} also gives
\eq{C_{,\phi}=\frac{U'(N_0)}{U'}C_{,\phi}(N_0).\label{eq-Sep-SumC}}
In the above equations, we have used primes on separate potentials as partial derivatives w.r.t the field, $N_0$ as $N$ at the boundary, and $U(X)$ as the value of $U$ at $N=X$ on a specific trajectory. After simple manipulation, we find \refeq{Sep-Sump} is exactly the same with the results obtained in Ref.~\cite{Vernizzi:2006ve}. For the second order derivatives, we can just take derivatives from \refeq{Sep-Sump}, while paying extra attention on $N_{0,\phi}$, which may be non-vanishing.

With the separable potential by sum \refeq{Sep-Sum}, if the energy densities of $\phi$ and $\psi$ dominate over other components, we are able to calculate the non-adiabaticity in energy density perturbations. Note that for the separable potentials, mentioned in \refsec{Two-field inflationary models}, the total energy density is dominated by the vacuum energy density of the waterfall field $\chi$, so the following discussion is not applicable to those cases.

For a separable potential, at the end of inflation we can define the non-adiabaticity perturbation in energy density, as
\eq{S\equiv\left.\frac{\delta\rho_\phi}{\rho_\phi}-\frac{\delta\rho_\psi}{\rho_\psi}\right|_e,}
in which $\rho_\phi$ is the energy density of $\phi$ and $\delta\rho_\phi$ denotes its perturbation. Here the subscript $e$ also means the end of inflation, i.e.\ taking the expressions at $N=N_0$. Under the slow roll approximations, we find
\eq{S=\left.\left(\frac{U'}{U}f_\phi-\frac{W'}{W}f_\psi\right)\right|_e(C_{,\phi}\delta\phi+C_{,\psi}\delta\psi).}
Remembering that $\zeta=N_{,\phi}\delta\phi+N_{,\psi}\delta\psi$, we obtain the power spectra of $S$ and cross-correlation between $S$ and $\zeta$, as
\eqa{P_S&\equiv&\langle S^2\rangle=\left.\left(\frac{U'}{U}f_\phi-\frac{W'}{W}f_\psi\right)^2\right|_e(C_{,\phi}^2+C_{,\psi}^2)\frac{H^2}{4\pi^2},\\
P_{S\zeta}&\equiv&\langle S\zeta\rangle=\left.\left(\frac{U'}{U}f_\phi-\frac{W'}{W}f_\psi\right)\right|_e(C_{,\phi}N_{,\phi}+C_{,\psi}N_{,\psi})\frac{H^2}{4\pi^2}.}
Their spectral tilts are given by:
\eqa{n_S-1&\equiv&-\frac{\dd\ln P_S}{\dd N}=-2\epsilon^2-2\frac{C_{,\phi}C_{,\phi}'+C_{,\psi}C_{,\psi}'}{C_{,\phi}^2+C_{,\psi}^2}\\
n_{S\zeta}-1&\equiv&-\frac{\dd\ln P_{S\zeta}}{\dd N}=-2\epsilon^2-\frac{C_{,\phi}N_{,\phi}'+C_{,\phi}'N_{,\phi}+C_{,\psi}N_{,\psi}'+C_{,\psi}'N_{,\psi}}{C_{,\phi}N_{,\phi}+C_{,\psi}N_{,\psi}}.}
Following Ref.~\cite{Komatsu:2008hk}, we define the non-adiabaticity parameter $\alpha$ and the cross-correlation parameter $\beta$ as
\eqa{\frac{\alpha}{1-\alpha}&\equiv&\frac{P_S}{P_\zeta}=\left.\left(\frac{U'}{U}f_\phi-\frac{W'}{W}f_\psi\right)^2\right|_e\frac{C_{,\phi}^2+C_{,\psi}^2}{N_{,\phi}^2+N_{,\psi}^2},\\
\beta&\equiv&-\frac{P_{S\zeta}}{\sqrt{P_SP_\zeta}}=-\frac{C_{,\phi}N_{,\phi}+C_{,\psi}N_{,\psi}}{\sqrt{(C_{,\phi}^2+C_{,\psi}^2)(N_{,\phi}^2+N_{,\psi}^2)}}.}
When comparing with the non-adiabaticity in the field perturbations, \refeq{CObs-alphat} and \refeq{CObs-betat}, we find that although the expressions of $\alpha$ and $\widetilde\alpha$ have their boundary parts different, $\beta$ and $\widetilde\beta$ only differ by a sign.

\ssecs{Separable by product}
For separable potentials by product, we use the generic form
\eq{V(\phi,\psi)=\mpl^4e^{U(\phi)+W(\psi)}.}
We can easily obtain:
\eq{N_{,\phi}=\frac{U'(N_0)}{U'(N)}N_{,\phi}(N_0).\label{eq-Sep-Prop}}
The second order derivatives can be derived in the same way from taking derivatives from \refeq{Sep-Prop}.

\bibliographystyle{jcap}
\bibliography{Main}
\end{document}